\documentclass[aps,preprint,nofootinbib]{revtex4}%
\usepackage{amsfonts}
\usepackage{amsmath}
\usepackage{amssymb}
\usepackage{float}
\usepackage{graphicx}%
\providecommand{\U}[1]{\protect\rule{.1in}{.1in}}

\begin{document}
 
\title{Lovelock black p-branes with fluxes}

\author{$^{1}$Adolfo Cisterna, $^{2}$Sebasti\'an Fuenzalida, and $^{2}$Julio Oliva}

\affiliation{$^{1}$Vicerrector\'ia Acad\'emica, Toesca 1783, Universidad Central de Chile, Santiago, Chile}

\affiliation{$^{2}$Departamento de F\'isica, Universidad de Concepci\'on, Casilla,
160-C, Concepci\'on, Chile.}

\email{adolfo.cisterna@ucentral.cl, sfuenzalida@udec.cl, juoliva@udec.cl}

\begin{abstract}
In this paper we construct compactifications of generic, higher curvature Lovelock theories of gravity over direct product spaces of the type $\mathcal{M}_D=\mathcal{M}_d \times \mathcal{S}^p $, with $D=d+p$ and $d\ge5$, where $\mathcal{S}^p$ represents an internal, Euclidean manifold of positive constant curvature. We show that this can be accomplished by including suitable non-minimally coupled $p-1$-form fields with a field strength proportional to the volume form of the internal space. We provide explicit details of this constructions for the Einstein-Gauss-Bonnet theory in $d+2$ and $d+3$ dimensions by using one and two-form fundamental fields, and provide as well the formulae that allows to construct the same family of compactification in any Lovelock theory from dimension $d+p$ to dimension $d$. These fluxed compactifications lead to an effective Lovelock theory on the compactfied manifold, allowing therefore to find, in the Einstein-Gauss-Bonnet case, black holes in the Boulware-Deser family.
\end{abstract}

\maketitle

\section{Introduction}

 In General Relativity (GR), Kaluza-Klein (KK) reductions on direct product spacetimes of the form $\mathcal{M}_4\times \mathcal{K}^{(D-4)}$, may be supported by fundamental $(p-1)-$forms $A_{[p-1]}$ \cite{Freund:1980xh}, where the field strength $F_{[p]}=dA_{[p-1]}$, is proportional to the volume form of the compact manifold $\mathcal{K}^{(D-4)}$. Abelian $p$-form fields are ubiquitous in fundamental theories, for example in eleven-dimensional supergravity there is a fundamental three-form $A_{[3]}$ whose dimensional reduction to ten dimensions leads to the $A_{[2]}$ and $A_{[3]}$ fields of IIA SUGRA. Also in the spectrum of Type IIB SUGRA there is a higher degree form, namely $A_{[4]}$ with a ``self-dual" field strength. On the other hand, in a four dimensional setup, a three-form $A_{[3]}$ allows to transform a fundamental cosmological parameter into an integration constant \cite{Henneaux:1984ji}, providing an explicit mechanism to explore, for example, the extended thermodynamics of black holes in Anti-de Sitter (AdS) \cite{Kastor:2009wy}-\cite{Dolan:2010ha}. Non-vanishing values of the $p-$forms are referred to as fluxes. All these theories acquire higher curvature corrections which may come from $\alpha'$ corrections when embedded in String Theory, or from the integration-out of massive degrees of freedom, leading to an effective theory. The precise form of such combinations, depend on the specific details of the fundamental theory and are defined up to the field redefinitions allowed by the intrinsic ambiguities of the perturbative approach. In this paper, as a toy model, we will consider particular combinations of higher curvature terms including also non-minimal couplings between matter and gravity \cite{Feng:2015sbw}. Such combinations (see below) lead to second order field equations and have the advantage of leading to explicit, analytic expressions for the black hole solution on the compactified spacetime.\\
Before constructing the new family of solutions, let us review the compactifications of the form $\mathcal{M}_4\times \mathcal{K}^{(D-4)}$, for GR in vacuum, in the presence of the Gauss-Bonnet combination, which is quadratic in the curvature. The action principle of Einstein-Gauss-Bonnet (EGB) gravity in arbitrary dimensions is given by
\begin{equation}
I_{EGB}[g]=\int{\sqrt{-g}d^Dx(R-2\Lambda+\alpha \mathcal{GB})}\ ,
\end{equation}
where $\alpha$ stands for the Gauss-Bonnet coupling which has mass dimension $-2$, and $\mathcal{GB}=R^2-4R_{AB}R^{AB}+R_{ABCD}R^{ABCD}$. To start, we focus on compactifications over direct product spaces of the type 
\begin{equation}
\mathcal{M}_{D}=\mathcal{M}_{d} \times \mathcal{K}^{2} \label{product}
\end{equation}
which translate into the $D=d+2-$dimensional spacetime metric
\begin{align}
ds^{2}&=g_{AB}dx^{A}dx^{B}=\tilde{g}_{\mu\nu}\left(y\right)dy^{\mu}dy^{\nu}+\hat{g}_{ij}\left(z\right)dz^{i}dz^{j}.\label{eq:line-element}
\end{align}
Here $\tilde{g}_{\mu\nu}\left(y\right)dy^{\mu}dy^{\nu}$ represents the $d-$dimensional spacetime manifold $\mathcal{M}_{d}$ and $\hat{g}_{ij}\left(z\right)dz^{i}dz^{j}$ stands for a $2-$dimensional Euclidean manifold $\mathcal{K}^{2}$. We demand on $\mathcal{K}^{2}$ to be of constant curvature, i.e. 
\begin{equation}
\hat{R}_{ijkl}=\gamma(\hat{g}_{ik}\hat{g}_{jl}-\hat{g}_{il}\hat{g}_{jk})
\end{equation}
where $\gamma$ defines the curvature radius $R_0=|\gamma|^{-1}$ of the internal manifold $\mathcal{K}^{2}$. Here after we use tilde and hat to denote objects that are intrinsically defined on $\mathcal{M}_d$ and $\mathcal{K}^p$, respectively.\\
Unlike GR with a cosmological term, direct product compactifications in EGB gravity on internal manifolds with non-vanishing constant curvature are possible even in vacuum. The new geometric term introduced in the action, the Gauss-Bonnet combination, removes the incompatibilities arising from the equations of motion on the brane and on the internal manifold, at the cost of both, fixing the aforementioned curvature to be negative and fixing the coupling constant $\alpha$ leading to a non-generic theory. Indeed, EGB equations are
\begin{equation}
R_{AB}-\frac{1}{2}g_{AB}R+\Lambda g_{AB}+\alpha H_{AB}=0\ , 
\end{equation}   
where
\begin{equation}
H_{AB}=2RR_{AB}-4R_{AC}R^{C}{}_{B}-4R^{CD}R_{ACBD}+2R_{ACDE}R_{B}{}^{CDE}-\frac{1}{2}g_{AB}\mathcal{GB}\ .\label{gb}
\end{equation}
is the Gauss-Bonnet tensor. By noticing that on the direct product spacetime (\ref{product}) the scalar curvature and the Gauss-Bonnet combination split as
\begin{align}
R&=\tilde{R}_d+2\gamma\ ,\\
\mathcal{GB}&=\tilde{\mathcal{GB}}_d+4\gamma \tilde{R}_d\ , 
\end{align}
with $d$ denoting quantities defined on $\mathcal{M}_{d}$, we observe that compatibility of the field equations \eqref{gb}, when projected along $\mathcal{M}_d$ and $\mathcal{K}^2$, imply that
\begin{align}
\gamma&=-\frac{1}{2\alpha(d-2)}\ , \label{cond1} \\
\Lambda&=-\frac{1}{8\alpha(d-2)}\ . \label{cond2}
\end{align}
The relation between the couplings $\alpha$ and $\Lambda$ leads to a non-generic theory, while $\alpha>0$ implies that $\gamma$ must be negative. Even more, compact Lobachevsky spaces $(\gamma<0)$ do not admit globally defined Killing vectors, therefore the Kaluza-Klein fluctuations of the metric around this solution, will not lead to well-defined Yang-Mills fields on the reduced spacetime.\\
	In order to obtain a positive curvature on the internal manifold a possibility is to dress it with minimally coupled $(p-1)-$form fields whose field strength can be naturally chosen as proportional to the volume form of such compact manifold, i.e. $F_{[p]}=dA_{[p-1]}\sim q_m \text{vol}(\mathcal{K})$. Although this approach is consistent when compactifying Einstein theory \cite{Freund:1980xh,RandjbarDaemi:1982hi}, it is not useful in more general gravitational theories as it is the case of EGB gravity since the presence of the $p-$form magnetic charge $q_m$, while removing the condition \eqref{cond2}, it will still lead to a negative curvature $\gamma$ for the internal space (see details below).\\
It will be then clear that in order to compactify EGB gravity in direct product spacetimes, on an internal manifold of positive constant curvature, we need to include non-minimal couplings between curvature tensors and $p-$form fields. For simplicity, we restrict to the family of combinations that lead to second order field equations and that fulfil the following two requirements:  First, the coupling between curvature tensors and the $p-$form fields should be quadratic in the field strength, ensuring that the equations of motion for the matter fields will be linear on the fundamental $p-$form. Secondly, the non-minimal couplings should be such that their contributions to the compatibility relations have the same number of the curvatures originally contained in the gravitational theory under consideration. For the case of EGB gravity, whose compatibility relation possesses a term proportional to the scalar curvature, what we need is a new term proportional to the Ricci scalar controlled now by the $p-$form magnetic charge. If this is not the case, for example, if the new term contains contributions that are quadratic in the curvature, then new incompatibilities may arise.\\
The paper is outlined as follows: In Section II we introduce our model, Lovelock gravity endowed with a non-minimally coupled antisymmetric field with field strength $F_{[p]}=dA_{[p-1]}$. 
Section III is devoted to compactify EGB theory non-minimally coupled to a $2-$form field. This first example illustrates how our model makes it possible to compactify a generic EGB theory on two-dimensional manifolds of positive constant curvature. Section IV extends this result to the case of three-dimensional internal manifolds of positive constant curvature by making use of non-minimally coupled $3-$form field strengths. For both cases we provide the explicit black hole solutions given by topological Boulware-Deser black 2-branes and black 3-branes, respectively. Section V offers a general analysis on how to use our model to compactify any Lovelock theory in arbitrary dimension by using non-minimally coupled $p-$forms. Some conclusions and further comments are given in section VI.

\section{The theory}
A natural theory suitable for compactifications of Lovelock gravity is the one proposed in \cite{Feng:2015sbw}, where the authors have constructed a higher-curvature theory of gravity with non-minimally coupled $p-$forms. The Lagrangian is built in terms of a polynomial invariant of the Riemann tensor and the field strength of Abelian gauge fields. Second order equations of motion are required for both, metric and matter fields avoiding the presence Ostrogradsky instabilities.
The construction is made in direct analogy with Lovelock theory, whose basic ingredients are the Euler integrands constructed from the full contraction of a generalized Kronecker deltas with Riemann tensors. In order to perform the non-minimal coupling with $p-$form fields, it is useful to introduce the following combination
\begin{equation}
Z^{A_1...A_p}{}_{B_1...B_p}=F^{A_1...A_p}F_{B_1...B_p}\ ,
\end{equation}
being $F_{[p]}=dA_{[p-1]}$, such that the Lagrangian is written as
\begin{equation}
\mathcal{L}^{(k,n)}_{p}=\frac{1}{2^k(p!)^{n}}\delta^{A_1\cdots A_{2k}C_1^{1}\cdots C_1^{p}\cdots C_n^{1}\cdots C_n^{p}}_{B_1\cdots B_{2k}D_1^{1}\cdots D_1^{p}\cdots D_n^{1}\cdots D_n^{p}}R^{B_1B_2}{}{}_{A_1A_2}\cdots R^{B_{2k-1}B_{2k}}{}{}_{A_{2k-1}A_{2k}}Z^{D_1^{1}\cdots D_1^{p}}{}_{C_1^{1}\cdots C_1^{p}}\cdots Z^{D_n^{1}\cdots D_n^{p}}{}_{C_n^{1}\cdots C_n^{p}}\ .\label{theory}
\end{equation}
Due to the Bianchi identity of the $p-$forms, only up to second derivatives of the fundamental fields will appear in the equations of motion \cite{Feng:2015sbw}. We will be concerned in particular with Lagrangians of the form (\ref{theory}) with $n=1$ therefore leading to a linear equation for the $(p-1)-$form potential. To gain familiarity with this structure, let us consider the first non-trivial low-lying term given by $m=n=1$, with $F_{AB}$ being the electromagnetic tensor, namely, 
\begin{equation}
\mathcal{L}^{(1,1)}_2=\frac{1}{4}\delta^{A_1A_2C_1C_2}_{B_1B_2D_1D_2}R^{B_1B_2}{}_{A_1A_2}Z^{D_1D_2}{}_{C_1C_2}=RF^2-4R_{AB}F^{AC}F^{B}{}_{C}+R_{ABCD}F^{AB}F^{CD}\ ,  \label{11theory}
\end{equation}
where we have explicitly expanded the Kronecker delta in order to illustrate the new contributions in the Lagrangian.
This is precisely the Horndeski electrodynamic introduced in \cite{Horndeski:1976gi}, where this theory was proposed as a natural non-minimally coupled electrodynamic that in the flat spacetime limit leads to Maxwell equations. Lagrangian (\ref{theory}) is then the natural generalization of (\ref{11theory}) to any higher power of the curvature and arbitrary $p-$form fields. As we will see in the next section, the combination (\ref{11theory}) is exactly what we need in order to compactify EGB theory on a $2$-dimensional internal manifold of positive constant curvature and we will moreover show that their higher dimensional extensions are naturally performed by going further in the degree of the considered $p-$form field\footnote{This is easily inferred from the fact that our field strengths are proportional to the volume form of the internal manifold.}.  

\section{Einstein-Gauss-Bonnet compactifications on $\mathcal{S}^2$}
Here we will show that in order to compactify EGB theory on an internal manifold of positive constant curvature we must include non-minimally coupled terms as the ones contained in (\ref{theory}). For compactifications over $\mathcal{S}^2$ the action principle reads
\begin{equation}
I_{d+2}\left[g,A_{[2]}\right]=\int{\sqrt{-g}d^{d+2}x[R-2\Lambda+\alpha\mathcal{GB}-\frac{1}{4}F_{AB}F^{AB}+\beta\mathcal{L}^{(1,1)}_2]}\ ,
\end{equation}
where the $2-$sphere is dressed with a Maxwell field proportional to the volume form of the internal manifold  $\sigma=\text{vol}(\mathcal{S}^2)$, namely
\begin{equation}
F_{ij}=q_m\sqrt{\hat{g}}\hat{\epsilon}_{ij}\ .  \label{2form}
\end{equation}
(see \eqref{eq:line-element}). Note that we have included the standard Maxwell kinetic term which is useful when considering a non-trivial bare cosmological constant \cite{RandjbarDaemi:1982hi}. The couplings $\alpha$ and $\beta$ have mass dimension $-2$. For the sake of concreteness we provide the field equations in an expanded form
\begin{equation}
R_{AB}-\frac{1}{2}g_{AB}R+\Lambda g_{AB}+\alpha H_{AB}=\frac{1}{2}F_{AC}F_{B}{}^{C}-\frac{1}{8}g_{AB}F^2+\beta T^{(1,1)}_{AB,2}, 
\end{equation}   
where $H_{AB}$ is given by (\ref{gb}) and
\begin{align}
T^{(1,1)}_{AB,2}=&R_{AB}F^2+2RF_{(A}^{\ \ C}F_{B)C}-8R_{D(A}F_{B)}^{\ C}F^{D}_{\ C}-4R_{CD}F_{\ (A}^CF_{\ B)}^D\nonumber\\  
& -3R_{CDE(A}F_{B)}^{\ \ E}F^{CD}-\frac{1}{2}g_{AB}\mathcal{L}^{(1,1)}_2+(g_{AB}\Box-\nabla_A\nabla_B)F^2\nonumber\\ 
&+4\nabla_C\nabla_{(A}(F_{B)D}F^{CD})-2\Box(F_{(A}^{\ \ C}F_{B)C})-2g_{AB}\nabla_{C}\nabla_{D}(F^{CE}F^{D}_{\ E})\nonumber\\  
& +2\nabla_C\nabla_D(F_{(A}^{\ \ C}F_{B)}^{\ \ D})\ ,
\end{align}
represents the contribution to the field equations of the term $\mathcal{L}^{(1,1)}_2$ defined in \eqref{theory}. After using the Bianchi identity for the field strength $F_{[2]}$, $T^{(1,1)}_{AB,2}$ can be cast in a manifestly second order fashion (for details see Section VI). The Maxwell equation is given by
\begin{equation}
\nabla_BF^{BA}+4\beta\nabla_B(RF^{AB}+4R^{C[A}F_{C}{}^{B]}+R^{AB}{}{}_{CD}F^{CD})=0\ ,
\end{equation}
which is automatically satisfied for configuration (\ref{2form}). For the $D=d+2-$dimensional product spacetime 
\begin{equation}
\mathcal{M}_{D}=\mathcal{M}_{d} \times \mathcal{S}^{2}\ ,
\end{equation}
the trace of the equations of motion on the brane and on the internal manifold yield 
\begin{align}
\frac{d-2}{2}(1+4\alpha\gamma+2\beta q_m^2)\tilde{R}_d+\frac{\alpha}{2}(d-4)\tilde{\mathcal{GB}}_d-\frac{d}{4}(4\Lambda-4\gamma+q_m^2)&=0\ ,\\
(2\beta q_m^2-1)\tilde{R}_d-\alpha\tilde{\mathcal{GB}}_d+2\Lambda-\frac{1}{2}q_m^2&=0\ ,
\end{align}
whose compatibility is ensured by 
\begin{align}
\gamma&=-\frac{\beta}{\alpha}\frac{(d-3)}{(d-2)}q_m^2-\frac{1}{2\alpha(d-2)}\ , \label{compgb01}\\
\Lambda&=-\frac{1}{4}\frac{\beta}{\alpha}\frac{d(d-3)}{(d-2)}q_m^2-\frac{1}{8}(d-2)q_m^2-\frac{d}{8\alpha(d-2)}\ . \label{compgb2}
\end{align}
From the previous relations it is direct to see that the values of the Gauss-Bonnet parameter as well as $\beta$ are completely free due to the presence of the magnetic charge $q_m$ and the curvature radius of the internal manifold $\mathcal{S}^{2}$. From Equation (\ref{compgb01}) we observe that the positivity of $\gamma$ implies a negative upper bound on $\beta$ and as we will see below, which is compatible with the existence of arbitrarily small black holes with positive entropy.\\
In order to integrate a specific solution let us take the 8-dimensional case, namely, $d=6$. The conditions (\ref{compgb01}) and \eqref{compgb2} fix the magnetic charge and the curvature of $\mathcal{S}^{2}$ respectively to
\begin{equation}
q_m^2=-\frac{3+16\alpha\Lambda}{2(4\alpha+9\beta)}, \ \ \gamma=\frac{12\beta\Lambda-1}{2(4\alpha+9\beta)}
\end{equation}
Then, assuming a Schwarzschild-like ansatz on the brane, Einstein equations lead to the following spacetime metric
\begin{equation}
ds^{2}=-F(r)dt^{2}+\frac{dr^{2}}{F(r)}+r^{2}  \frac{dz_{1}d\bar{z}_{1}%
+dz_{2}d\bar{z}_{2}}{ \left(1+\frac{K}{4}\left(  z_{1}\bar{z}_{1}+z_{2}\bar{z}%
_{2}\right) \right)  ^{2}}+\frac{dwd\bar{w}}{(1+\frac{\gamma}{4}w\bar{w})^2}%
\end{equation}
where
\begin{equation}
F\left(  r\right)  =K+\frac{\alpha+\beta\left(  3+4\Lambda\alpha\right)
}{6\alpha\left(  9\beta+4\alpha\right)  }r^{2}\left[  1-\sqrt{1+\frac
{3\alpha\left(  9\beta+4\alpha\right)  \left(  16\Lambda\alpha+24\Lambda
\beta+1\right)  }{40(\alpha+\beta\left(  3+4\Lambda\alpha\right)  )}-\frac
{\mu}{r^{5}}}\right]\ ,\label{F6}
\end{equation}
which describes a black 2-brane in eight dimensions constructed out from the direct product of a six-dimensional topological Boulware-Deser black hole \cite{Mann:1996gj,Boulware:1985wk,Cai:2001dz} lying on the transverse section and a two-dimensional sphere. Here $\mu$ is an integration constant. It is interesting to notice that for a particular value of non-minimal matter coupling $\beta$, the second term below the square root in equation \eqref{F6} can vanish, leading to an asymptotically flat black hole on the brane. This solution is obtained by algebraically solving the corresponding Wheeler's polynomial \cite{Wheeler:1985nh,Wheeler:1985qd} that arises from the direct integration of
\begin{equation}
\frac{d}{dr}\left[12\alpha r(K-F(r))^2+r^3(1-2q_m^2\beta)(K-F(r))+\frac{r^5}{80\alpha}((4\alpha+6\beta)q_m^2+1)\right]=0 \label{wheeler1}
\end{equation}
on which the compatibility relations have been already considered, ensuring that not only Einstein equations on $\mathcal{M}_6$ are fulfilled, but also those along $\mathcal{S}^2$. Notice that equation (\ref{compgb01}) requires $\beta<0$, implying that the first two terms in (\ref{wheeler1}) have the same sign. Comparing this with the six-dimensional Wheeler polynomial of Einstein-Gauss-Bonnet gravity (see e.g Eq (5.8) of \cite{Maeda:2011ii}) we observe that the negativity of $\beta$ implies the existence of arbitrarily small, spherically symmetric black holes with positive entropy on the brane.
 
\section{Einstein-Gauss-Bonnet compactifications on $\mathcal{S}^3$}
Following a similar strategy as the one of the previous section, in order to compactify EGB gravity over $\mathcal{S}^3$ we will dress the internal manifold with a $3-$form field strength $H_{[3]}$. Our action is then given by
\begin{equation}
I_{d+3}\left[g,A_{[3]}\right]=\int{\sqrt{-g}d^{d+3}x[R-2\Lambda+\alpha\mathcal{GB}-\frac{1}{6}H_{ABC}H^{ABC}+\beta\mathcal{L}^{(1,1)}_3]}\ ,
\end{equation}
where
\begin{equation}
\mathcal{L}^{(1,1)}_3=\frac{1}{12}\delta_{B_1B_2D_1D_2D_3}^{A_1A_2C_1C_2C_3}R^{B_1B_2}{}{}_{A_1A_2}Z^{D_1D_2D_3}{}_{C_1C_2C_3}\ .
\end{equation}
The equations of motion are given by
\begin{equation}
R_{AB}-\frac{1}{2}g_{AB}R+\Lambda g_{AB}+\alpha H_{AB}=\frac{1}{2}H_{A}{}^{CD}H_{BCD}-\frac{1}{12}g_{AB}H^2+\beta T^{(1,1)}_{AB,3}\ ,
\end{equation}
with the energy-momentum tensor of the non-minimal coupling term defined as 
\begin{align}
T^{(1,1)}_{AB,3}=&\frac{1}{24}g_{AB}\delta^{A_1A_2C_1C_2C_3}_{B_1B_2D_1D_2D_3}R^{B_1B_2}{}{}_{A_1A_2}Z^{D_1D_2D_3}{}_{C_1C_2C_3}-\frac{1}{12}\delta^{A_1A_2C_1C_2C_3}_{B_1\left(A\right|D_1D_2D_3}R^{B_1}{}_{\left|B\right)A_1A_2}Z^{D_1D_2D_3}{}_{C_1C_2C_3}\nonumber\\
&+\frac{1}{6}\delta^{A_1A_2C_1C_2C_3}_{\left(A\right|B_2D_1D_2D_3}g_{A_2\left|B\right)}\nabla_{A_1}F^{D_1D_2D_3}\nabla^{B_2}F_{C_1C_2C_3}+\frac{1}{2}\delta^{A_1A_2C_1C_2C_3}_{\left(A\right|B_2D_1D_2D_3}g_{A_2\left|B\right)}R^{D_1}{}_{E}{}^{B_2}{}_{A_1}Z^{ED_2D_3}{}_{C_1C_2C_3}\nonumber\\
&-\frac{1}{4}\delta^{A_1A_2C_1C_2C_3}_{B_1B_2\left(A\right|D_2D_3}R^{B_1B_2}{}_{A_1A_2}Z_{\left|B\right)}{}^{D_2D_3}{}_{C_1C_2C_3}\ . 
\end{align}
Here we have used Bianchi identities in order to cast the field equations in a manifestly second order form. We observe that the trace of the equations of motion on the brane and on the internal manifold respectively lead to
\begin{align}
\frac{d-2}{2}(1+12\alpha\gamma+6\beta q_m^2)\tilde{R}_d+\frac{\alpha}{2}(d-4)\tilde{\mathcal{GB}}-\frac{d}{2}(2\Lambda-6\gamma+q_m^2)&=0,\\
\left(9\beta q_m^2-6\alpha\gamma-\frac{3}{2}\right)\tilde{R}_d-\frac{3}{2}\alpha \tilde{\mathcal{GB}}_d+3\Lambda-3\gamma-\frac{3}{2}q_m^2&=0 \ .
\end{align}
Now, the compatibility will be ensured provided
\begin{align}
\gamma&=-\frac{3\beta}{2\alpha}\frac{(d-3)}{(d-1)}q_m^2-\frac{1}{4\alpha(d-1)},\\
\Lambda&=-\frac{3\beta}{4\alpha}\frac{(d+2)(d-3)}{d-1}q_m^2-\frac{1}{4}(d-2)q_m^2-\frac{1}{8\alpha}\frac{(d+2)}{(d-1)}\ .
\end{align}
As in the previous section, a positive value of $\gamma$ implies a negative upper bound on $\beta$. The magnetic charge and the $\gamma$ are then fixed by
\begin{equation}
q_m^2=-\frac{1+5\alpha\Lambda}{5\alpha+18\beta}, \  \gamma=-\frac{9}{10}\frac{\beta q_m^2}{\alpha}-\frac{1}{20\alpha}
\end{equation}
Taking as an example the $9-$dimensional case ($d=6$), and a Schwarzschild-like ansatz on the brane, namely
\begin{equation}
ds^2=-F(r)dt^2+\frac{dr^2}{F(r)}+\frac{r^2(dy_1^2+dy_2^2+dy_3^2+dy_4^2)}{(1+\frac{K}{4}(y_1^2+y_2^2+y_3^2+y_4^2))^2}+\frac{(dz_1^2+dz_2^2+dz_3^2)}{(1+\frac{\gamma}{4}(z_1^2+z_2^2+z_3^2))^2}\ ,
\end{equation}
leads to the following lapse function
\begin{equation}
F\left(  r\right)  =\gamma+\frac{\alpha+6\beta\left(  1+2\alpha\Lambda\right)
}{6\left(  5\alpha+18\beta\right)  \alpha}r^{2}\left[  1-\sqrt{1+\frac
{3\alpha\left(  5\alpha+18\beta\right)  \left(  10\Lambda\alpha+18\beta
\Lambda+1\right)  }{20}-\frac{\mu}{r^{5}}}\right] \ .
\end{equation}
This solution represents a black string in $9-$dimensions constructed with the direct product of a $6-$dimensional, topological Boulware-Deser black hole and a $3-$dimensional internal manifold of positive constant curvature. It is obtained by direct integration of the associated Wheeler's polynomial
\begin{equation}
\frac{d}{dr}\left[12\alpha r(K-F(r))^2+\frac{4}{5}r^3(1-12\beta q_m^2)(K-F(r))+\frac{r^5}{100\alpha}(1+10\alpha q_m^2+18q_m^2\beta)\right]=0 \ ,
\end{equation}
which leads to the integration constant $\mu$. 
The same conditions on $\beta$ as those given in the previous section, lead to well behaved black holes of arbitrary small radius. In the next section we provide the general formulae that allows to reduce a Lovelock theory from $D=d+p$ to dimension $d\ge5$ for arbitrary values of the couplings. 

\section{General solutions for arbitrary Lovelock theories}

To extend the previous results to arbitrary Lovelock theories, in dimension $D=d+p$, we consider the full Lagrangian (\ref{theory}) with $n=1$, leading to the following action functional  
\begin{align}
I_{d+p}\left[g,A_{[p]}\right]=&\int\sqrt{-g}d^{d+p}x\left[\sum^{N+1}_{k=0}\frac{\alpha_{k}}{2^{k}}\delta^{A_1\cdots A_{2k}}_{B_1\cdots B_{2k}}R^{B_1B_2}{}{}_{A_{1}A_{2}}\cdots R^{B_{2k-1}B_{2k}}{}{}_{A_{2k-1}A_{2k}}\right.\nonumber\\
&\left.+\sum^{N}_{k=0}\frac{\beta_{k}}{2^{k}p!}\delta^{A_1\cdots A_{2k}C_1\cdots C_p}_{B_1\cdots B_{2k}D_1\cdots D_p}R^{B_1B_2}{}{}_{A_{1}A_{2}}\cdots R^{B_{2k-1}B_{2k}}{}{}_{A_{2k-1}A_{2k}}Z_{(k)}^{D_1\cdots D_p}{}_{C_1\cdots C_p}\right]\ .\label{ec:action-p}
\end{align}
Even more we have extended the theory by considering $k$, in principle different, $p-1$ fundamental forms, with field strengths proportional to the volume form of the internal manifold. The sub-index $k$ on the bilinears $Z_{(k)}$ runs from $1$ to $N$. These are required to achieve compatibility for each geometrical term included in the Lovelock Lagrangian, maintaining generic gravitational couplings $\alpha_k$\footnote{For the $p=1$ case this accommodates the axionic black strings constructed in \cite{Cisterna:2018mww}.}. Note that in the previous sections, since only three Lovelock terms were present, we were able to achieve compatibility for arbitrary values of the couplings including a single $A_{[p-1]}$ magnetic configuration.
Varying the action (\ref{ec:action-p}) with respect to the metric one obtains the field equations
\begin{align}
\sum^{N+1}_{k=0}\alpha_{k}E^{\left(k\right)}_{AB}&=\sum^{N}_{k=0}\beta_{k}T^{\left(k,1\right)}_{AB,p}\ , \label{ec:field-equations-p}
\end{align}
where $E^{\left(k\right)}_{AB}$ and $T^{\left(k,1\right)}_{AB,p}$ are the Lovelock tensor and the energy-momentum tensor of the fundamental $(p-1)-$form, respectively defined as
\begin{align}
E^{\left(k\right)}_{AB}&=-\frac{1}{2^{k+1}}g_{\left(A\right|C}\delta^{CA_{1}\cdots A_{2k}}_{\left|B\right)B_1\cdots B_{2k}}R^{B_1B_2}{}{}_{A_{1}A_{2}}\cdots R^{B_{2k-1}B_{2k}}{}{}_{A_{2k-1}A_{2k}}\ ,\label{ec:lovelock-tensor-p}
\end{align}
and
\begin{align}
T^{\left(k,1\right)}_{AB,p}=&\frac{1}{2^{k+1}p!}g_{AB}\delta^{A_1\cdots A_{2k}C_1\cdots C_p}_{B_1\cdots B_{2k}D_1\cdots D_p}R^{B_1B_2}{}{}_{A_{1}A_{2}}\cdots R^{B_{2k-1}B_{2k}}{}{}_{A_{2k-1}A_{2k}}Z_{(k)}^{D_1\cdots D_p}{}_{C_1\cdots C_p}\nonumber\\
&-\frac{k}{2^{k}p!}\delta^{A_1A_2\cdots A_{2k}C_1\cdots C_p}_{B_1\left(A\right|\cdots B_{2k}D_1\cdots D_p}R^{B_1}{}_{\left|B\right)A_1A_2}R^{B_3B_4}{}{}_{A_3A_4}\cdots R^{B_{2k-1}B_{2k}}{}{}_{A_{2k-1}A_{2k}}Z_{(k)}^{D_1\cdots D_p}{}_{C_1\cdots C_p}\nonumber\\
&+\frac{2k}{2^{k}p!}\delta^{A_1\cdots A_{2k}C_1\cdots C_p}_{\left(A\right|\cdots B_{2k}D_1\cdots D_p}g_{A_2\left|B\right)}R^{B_3B_4}{}{}_{A_3A_4}\cdots R^{B_{2k-1}B_{2k}}{}{}_{A_{2k-1}A_{2k}}\nabla_{A_1}F_{(k)}^{D_1\cdots D_p}\nabla^{B_2}F_{(k)}{}_{C_1\cdots C_p}\nonumber\\
&+\frac{2pk}{2^{k}p!}\delta^{A_1\cdots A_{2k}C_1\cdots C_p}_{\left(A\right|\cdots B_{2k}D_1\cdots D_p}g_{A_2\left|B\right)}R^{B_3B_4}{}{}_{A_3A_4}\cdots R^{B_{2k-1}B_{2k}}{}{}_{A_{2k-1}A_{2k}}R^{D_1}{}_{E}{}^{B_2}{}_{A_1}Z_{(k)}^{ED_2\cdots D_p}{}_{C_1\cdots C_p}\nonumber\\
&-\frac{p}{2^{k}p!}\delta^{A_1\cdots A_{2k}C_1\cdots C_p}_{B_1\cdots B_{2k}\left(A\right|\cdots D_p}R^{B_1B_2}{}{}_{A_{1}A_{2}}\cdots R^{B_{2k-1}B_{2k}}{}{}_{A_{2k-1}A_{2k}}Z_{(k)}{}_{\left|B\right)}{}^{D_2\cdots D_p}{}_{C_1\cdots C_p}\ .\label{ec:energy-momentum-tensor-k-p}
\end{align}
Notice that the field equations are written in a manifestly second order fashion.
As before, we shall consider a direct product spacetime $\mathcal{M}_{D}=\mathcal{M}_d\times\mathcal{S}^p$, which translates into the $D=d+p-$dimensional spacetime metric
\begin{align}
ds^{2}&=g_{AB}dx^{A}dx^{B}=\tilde{g}_{\mu\nu}\left(y\right)dy^{\mu}dy^{\nu}+\hat{g}_{ij}\left(z\right)dz^{i}dz^{j}\ ,\label{ec:metric-ansatz-p}
\end{align}
where the explicit form of $\hat{g}_{ij}\left(z\right)$ is represented by   
\begin{align}
\hat{g}_{ij}dz^{i}dz^{j}&=\frac{d\vec{z}\cdot d\vec{z}}{\left(1+\frac{\gamma}{4}\vec{z}^{2}\right)^{2}}\ ,\label{ec:extented-part-metric-p}
\end{align}
with $\gamma$ defining the curvature radius $R_0=|\gamma|^{-1}$. Mimicking the particular case of Einstein-Gauss-Bonnet theory in $d+2$ and $d+3$ dimensions we use the following ansatz for the field strengths, $F_{(k),[p]}\sim q_{m,k}^2 Vol(\mathcal{S}^p)$, namely
\begin{align}
F_{(k)}{}_{i_1\cdots i_p}&=\frac{q_{m,k}}{\left(1+\frac{\gamma}{4}\vec{z}^{2}\right)^{p}}\hat{\epsilon}_{i_1\cdots i_p}\ .\label{ec:ansatz-F-p}
\end{align}
To obtain the trace of the field equations along the $\mathcal{M}_d$ dimensional brane and the internal manifold $\mathcal{S}^p$, for simplicity, we separately analyze the Lovelock tensor (\ref{ec:lovelock-tensor-p}) and the energy-momentum tensor (\ref{ec:energy-momentum-tensor-k-p}). Looking at the trace of the components of the Lovelock tensors (\ref{ec:lovelock-tensor-p}) on the brane, we obtain that 
\begin{align}
\tilde{g}^{\mu\nu}E^{\left(k\right)}_{\mu\nu}&=-\frac{1}{2}\sum^{k}_{j=0}\binom{k}{k-j}\frac{p!}{(p+2j-2k)!}\left(d-2j\right)\gamma^{k-j}\tilde{\mathcal{L}}^{(j)}_{d}\ ,\label{ec:trace-lovelock-tensor-1-p}
\end{align}
where the $j$-th Lovelock Lagrangian evaluated on the brane is
\begin{align}
\tilde{\mathcal{L}}^{(j)}_{d}&=\frac{1}{2^{j}}\delta^{\mu_1\cdots\mu_{2j}}_{\nu_{1}\cdots\nu_{2j}}\tilde{R}^{\nu_1\nu_2}{}{}_{\mu_1\mu_2}\cdots \tilde{R}^{\nu_{2j-1}\nu_{2j}}{}{}_{\mu_{2j-1}\mu_{2j}}\ .
\end{align}
On the other hand, looking at the $(i,j)$ components of the Lovelock tensors and taking the partial trace we have
\begin{align}
\hat{g}^{ij}E^{\left(k\right)}_{ij}&=-\frac{1}{2}\sum^{k}_{j=0}\binom{k}{k-j}\frac{p!}{\left(p+2j-2k-1\right)!}\gamma^{k-j}\tilde{\mathcal{L}}^{(j)}_{d}\ .\label{ec:trace-lovelock-tensor-2-p}
\end{align}
Performing the same computations for the energy-momentum tensor (\ref{ec:energy-momentum-tensor-k-p}) we obtain 
\begin{align}
\tilde{g}^{\mu\nu}T^{\left(k,1\right)}_{\mu\nu,p}&=\frac{p!}{2}\left(d-2k\right)q^{2}_{m,k}\tilde{\mathcal{L}}^{(k)}_{d}\ ,\label{ec:trace-energy-momentum-tensor-1-p}
\end{align}
and
\begin{align}
\hat{g}^{ij}T^{\left(k,1\right)}_{ij,p}&=-\frac{p^{2}}{2}\left(p-1\right)!q^{2}_{m,k}\tilde{\mathcal{L}}^{(k)}_{d}\ .\label{ec:trace-energy-momentum-tensor-2-p}
\end{align}
Therefore, the partial traces of the field equations (\ref{ec:field-equations-p}) lead to the following two, scalar constraints over $\mathcal{M}_d$
\begin{align}
&\sum^{N+1}_{k=0}\sum^{k}_{j=0}\binom{k}{k-j}\frac{p!}{(p+2j-2k)!}\left(d-2j\right)\gamma^{k-j}\alpha_{k}\tilde{\mathcal{L}}^{(j)}_{d}+\sum^{N+1}_{k=0}p!\left(d-2k\right)q^{2}_{m,k}\beta_{k}\tilde{\mathcal{L}}^{(k)}_{d}=0\ ,\label{ec:final-field-eq-p-mu-mu}\\
&\sum^{N+1}_{k=0}\sum^{k}_{j=0}\binom{k}{k-j}\frac{p!}{(p+2j-2k-1)!}\gamma^{k-j}\alpha_{k}\tilde{\mathcal{L}}^{(j)}_{d}-\sum^{N+1}_{k=0}p^{2}\left(p-1\right)!q^{2}_{m,k}\beta_{k}\tilde{\mathcal{L}}^{(k)}_{d}=0\ ,\label{ec:final-field-eq-p-i-i}
\end{align}
where we have assumed that $\beta_{N+1}=0$.\\
Therefore, from (\ref{ec:final-field-eq-p-mu-mu})-(\ref{ec:final-field-eq-p-i-i}) we must impose compatibility conditions which fix the integration constants $q_{m,k}$ given in the ansatz (\ref{ec:ansatz-F-p}), as well as the curvature $\gamma$ of the internal manifold. Once the compatibility relations are explicitly computed, generic Einstein-Lovelock field equations will determine the metric on the reduced spacetime $\mathcal{M}_d$. Assuming spherical symmetry on the later manifold, leads to an effective Wheeler's polynomial, providing black $p-$branes with fluxes, by algebraic integration. 

\section{Further comments}

Here, we have tackled the problem of compactifications of Einstein-Gauss-Bonnet gravity on direct product spaces of the form
\begin{equation}
\mathcal{M}_D=\mathcal{M}_d \times \mathcal{S}^p
\end{equation}
where $d\ge5$ and $\mathcal{S}^p$ represents an internal manifolds of positive constant curvature $\gamma$ defining the curvature radius $R_0=|\gamma|^{-1}$, the latter being therefore equivalent to a smooth quotient of the round sphere $S^p$. This family of spacetimes have been explored in different setups in Lovelock theories in vacuum. For example, black $p-$branes in theories containing a single Lovelock term were constructed in \cite{Giribet:2006ec}, and asymptotically AdS black holes on the brane in Lovelock theories with a single AdS-vacuum were reported in \cite{Kastor:2006vw}. In the presence of a non-trivial torsion, compactifications of EGB theory have been performed in \cite{Corralon}, black hole solutions in dimension eight were constructed in \cite{Canfora:2008ka}, while references \cite{Pons:2014oya} and \cite{Dadhich:2015nua} provided the construction of solutions given by products of spheres. A general analysis on the existence of neutral black branes in Lovelock theories was given in \cite{Kastor:2017knv}. As we have reviewed, in vacuum, self-consistent compactifications of EGB theory forces the curvature of the internal manifold to be negative, and also impose a fine-tuning between the Gauss-Bonnet couplings to cosmological term in the action. We showed that by including minimally coupled magnetic forms $F_{[p]}=dA_{[p-1]}$ we only obtain a modification of the relation between $\alpha$ and $\Lambda$, but still getting a negative value for $\gamma$. We have shown that, in order to achieve a positive curvature on the internal manifold, suitable non-minimal couplings between the curvature tensor and the fundamental fields $A_{[p-1]}$, as the one defined in \cite{Feng:2015sbw}, must be included. These non-minimal couplings lead to second order field equations for both the metric and the Abelian gauge fields. In fact, by considering these couplings we have successfully eliminated the fine-tuning of the Gauss-Bonnet coupling $\alpha$ in terms of the cosmological term $\Lambda$ in the action, while at the same time we have obtained a positive value for the curvature of the internal manifold. Both conditions are possible due to the presence of an extra parameter in the solution, namely the integration constant accounting for the magnetic charge of the non-minimally coupled Abelian gauge field $A_{[p-1]}$.\\
We have explicitly performed the compactifications of EGB theory in the case $D=d+2$ and $D=d+3$ by using $2-$ form and $3-$form fields coupled with the Riemann tensor according to the lines of \cite{Feng:2015sbw}. We have considered in particular the curved branes constructed with topological Boulware-Deser black holes on the branes. The former can also be asymptotically flat. We have provided as well the general formulae for the compactification of any Lovelock theory in dimension $D=d+p$ to dimension $d\geq 5$ by giving the explicit form of the traces of the field equations on the brane and along the internal manifold. When more than three Lovelock terms are included in the action, one needs to consider at least two, in principle different non-minimally coupled gauge field, if one is interested in achieving the compactifications for generic values of the Lovelock couplings.\\
Exact solutions in Lovelock theories supported by $p-$forms have been explored in \cite{Bardoux:2010sq} for black holes in EGB theory and also beyond the quadratic theory, on $R^4$ string inspired models in ten dimensions \cite{Giribet:2018hnl}. Also, $p-$forms lead to the construction of charged black strings with planar extended dimensions in theories containing a single Lovelock term \cite{Giacomini:2018sho}.\\
Finally, we comment on interesting generalizations that this work suggests. When working with these kind of theories containing terms with higher power of the curvature in the Lovelock family and consequently with higher dimensions, it is suggesting to explore for a self-contained procedure to make contact with Einstein theory in dimension four. This point has been partially addressed in \cite{Canfora:2008iu}, where the authors show that Lovelock gravity can be consistently compactified to four dimensional Einstein gravity with a positive cosmological constant starting from the cubic Lovelock theory, at the cost of introducing a fine-tuning between the couplings that is not protected by any symmetry. Preliminary results using our approach based on (\ref{theory}), suggest that it is possible to compactify six dimensional EGB theory to four dimensional Einstein theory, without requiring any fine-tuning within the coupling constants of the theory. Work along these lines is in progress \cite{soon1}.

\section{Acknowlegdments}
We thank the insights of Marcela Lagos as well as her contribution to early stages of this work. We also thank Fabrizio Canfora for discussions and enlightening comments. A. C. is supported by Fondo Nacional de Desarrollo
Cient\'{\i}fico y Tecnol\'{o}gico Grant No. 11170274 and Proyecto Interno Ucen I+D-2018, CIP2018020. The research of J.O. is supported
in part by the Fondecyt Grant 1181047.

\end{document}